\documentclass[prb,aps,twocolumn,floats,showpacs]{revtex4}
\usepackage{amsmath,amssymb}

\usepackage{epsfig}

\newcommand{\be}{\begin{equation}}
\newcommand{\ee}{\end{equation}}
\newcommand{\bea}{\begin{eqnarray}}
\newcommand{\eea}{\end{eqnarray}}

\DeclareMathOperator{\Sign}{sgn}

\DeclareMathOperator{\re}{Re}

\begin{document}
\title{Momentum distribution and tunneling density of states of one-dimensional Fermionic $SU(N)$ Hubbard model}
\author{Shuang Liang$^1$}
\author{Deping Zhang$^2$}
\author{Wei Chen$^{1}$}
\email{chenweiphy@nju.edu.cn}
\affiliation{$^1$Department of Physics, Nanjing University and National Laboratory of Solid State Microstructures, Nanjing 210093, China}
\affiliation{$^2$Nuctech Company Limited, Beijing 100084,  China}

\begin{abstract}
   We study the one-dimensional Fermionic Hubbard model with $SU(N)$ spin symmetry in the incommensurate filling case. The basic properties of Green's function, momentum distribution and tunneling density of states of the system at low temperature are studied in the frame work of Luttinger liquid theory combined with Bethe Ansatz solutions for arbitrary interaction. In the strong interacting case, the system enters the spin-incoherent regime at intermediate temperature $E_{{\rm spin}}<T\ll E_c$ and we obtain the Green's function and tunneling density of states by generalizing the path integral approach for the $SU(2)$ case to the $SU(N)$ case in this regime. The theoretical results we obtained agree qualitatively with the experiments on the one-dimensional alkaline earth atomic system with $SU(N)$ spin symmetry. The similarities and difference between the one-dimensional $SU(N)$ Fermionic Hubbard system at large $N$ and the one-dimensional spinless Bosonic system are also investigated.
\end{abstract}


\maketitle

\section{Introduction}
Correlations have been one of the most important topics in condensed matter physics in recent decades and have resulted in a large number of  exotic and fascinating phenomena. The correlation effects strongly depend on dimensionality and become more significant in lower dimensions in general. One remarkable  example is the breakdown of the famous Fermi liquid theory in one dimension~\cite{Giamarchi2004}. The correlation in one dimension results in collective excitations in interacting systems and the 1D system has to be described by the Luttinger liquid theory, according to which a series of exotic phenomena emerge, such as charge-spin separation, collective excitations, non-universal power law correlation etc~\cite{Giamarchi2004}. 

While the one-dimensional Fermionic systems with two spin components have been widely studied both theoretically and experimentally~\cite{Giamarchi2004, Yang1967, Sutherland1968, Voit1995, Fiete2004, Fiete2007, Cheianov2004, Cheianov2005,Yacoby1996, Bockrath1999}, the one-dimensional Fermionic system with large spin and $SU(N)$ symmetry has been realized only in recently years in cold atom systems with alkaline-earth atoms~\cite{Gazalilla2009, Gorshkov2010, Taie2012, Fallani2014}. Since the electronic angular momentum of the ground state of alkaline-earth atoms is zero, there is no coupling between the nuclear spin $I=9/2$ and the electron angular momentum for such state. For the reason, the s-wave scattering of the alkaline-earth atoms is independent of the nuclear spin, which results in an $SU(N)$ spin symmetry of the  alkaline-earth atom systems where $N=2I+1$ can be as large as $10$~\cite{Gorshkov2010, Fallani2014}. Such system is proposed to simulate a number of models with $SU(N)$ symmetry, such as the Kugel-Khomskii model, the Kondo lattice model, the Heisenberg and Hubbard model~\cite{Gorshkov2010, Rey2010, Gorshkov2016}.

 By trapping the alkaline-earth atoms into a one-dimensional optical lattice and applying optical spin manipulation technique, a one-dimensional Fermionic system with $SU(N)$ symmetry with $N$ varying from $1$ to $10$ was realized in experiments~\cite{Fallani2014}. The role of the spin multiplicity in the correlated one-dimensional system was investigated by measuring the momentum distribution and spectrum function for different $N$. The experiments reveal interesting properties for the system with large $N$, e.g., with the increase of $N$, the physical properties of the one- dimensional Fermionic system approach the behavior of one-dimensional Bosonic systems~\cite{Fallani2014, Yang2011}.

On the theoretical side, a number of  sophisticated techniques, such as Luttinger liquid theory~\cite{Schulz1990, Assaraf1999, Giamarchi2004, Fiete2004, Fiete2007}, conformal field theory~\cite{Izergin1989, Frahm1990, Frahm1993}, Bethe Ansatz~\cite{Haldane1981, Frahm1990, Frahm1993, Guan2012}, quantum inverse-scattering technique (QIST) etc.~\cite{Berkovich1987, Berkovich1991, Izergin1998}, have been employed to study the physical properties of one-dimensional Hubbard model. However, each of these techniques reveals only one face of the system and is applicable only in certain conditions, e.g., the Luttinger liquid theory is valid only for low energy physics and small interaction, the conformal field theory is applicable only in conformal invariant case, the solution of Bethe Ansatz and quantum inverse-scattering technique is often too involved to extract physical information and only very limited cases are solvable by these two methods. For the reason, a complete and consistent picture of the basic properties of the one-dimensional $SU(N)$ Hubbard model, such as the correlation function, the spectrum and momentum distribution etc, at arbitrary interaction and different temperature regimes is still lacking, especially for the $N>2$ case.

 In this work, we study the basic properties of Green's function, momentum distribution and tunneling density of states of the one-dimensional Fermionic $SU(N)$ Hubbard model mainly in the framework  of Luttinger liquid theory combined with Bethe Ansatz solutions, and discuss supplemental approaches in the regime when these two methods are not available. We focus on the incommensurate filling case, i.e., $\nu\neq p/q$ where $p$ and $q$ are coprime. In this case, the Umklapp processes are absent and the back scatterings of the Hubbard interaction are irrelevant. The system then stays in the metallic phase~\cite{Capponi2016} and the low energy physics of the one-dimensional interacting system can be described by Luttinger liquid theory. The properties of the system at low temperature are determined by the Luttinger parameters $K_\alpha$ which depends on the interaction. For weak interaction, the Luttinger parameters $K_\alpha$ can be obtained in the continuum limit from the Luttinger liquid theory. For strong interaction, the Luttinger liquid theory is no longer valid and we extract $K_\alpha$ from the Bethe Ansatz solutions for arbitrary interaction. The two methods together  give a complete description of the system at low temperature $T\ll E_{{\rm spin}}$. The theoretical results of the momentum distribution we obtained in this regime agree qualitatively with the experiments~\cite{Fallani2014}.

 In the strong interacting case, the charge and spin modes separate significantly and the system enters an intermediate spin-incoherent regime as the temperature increases to $E_{{\rm spin}}\ll T\ll E_c$. The charge mode in this regime can still be described by a spinless Luttinger liquid, however, the spin-part is disordered and the Luttinger liquid description is no longer valid. 
 We then generalize the path integral approach of computing the Green's function  for the $SU(2)$ case in this regime  in Ref.~\cite{Fiete2004, Fiete2007}  to the $SU(N)$ case in this work. The Green's function of the system decays exponentially in the spin-incoherent regime instead of a power law in the Luttinger liquid regime and the tunneling density of states diverges at low energy in the former case instead of vanishing as in the latter.
 The momentum distribution in the spin-incoherent regime is however not available from the path integral method in a general case. We then briefly discuss the special situation of infinite repulsion for which the density matrix and so the momentum distribution is exactly solvable from the quantum inverse-scattering technique (QIST) in the spin-incoherent regime~\cite{Cheianov2004, Cheianov2005, Berkovich1987, Berkovich1991, Izergin1998}. 

 We also discussed the similarity and difference between the one dimensional $SU(N)$ Fermionic system with large $N$ and the one-dimensional spinless Bosonic system.
Though it was discovered that the ground state energy of the two systems at $N\to \infty$ is the same at the same interaction and particle density~\cite{Yang2011}, indicating Bosonic behavior for the $SU(N)$ Fermionic system at $N\to \infty$~\cite{Fallani2014}, we show that the single particle Green's function and momentum distribution of the two systems are different due to the different statistics they obey.

 This paper is organized as follows: In section \ref{sec:model}, we review the one-dimensional $SU(N)$ Hubbard model and the single particle Green's function obtained in the Luttinger liquid regime. In section \ref{sec:LL_parameters}, we obtain the Luttinger liquid parameter from the solution of Bethe Ansatz for arbitrary interactions. In section \ref{sec:momemtum_distribution}, we calculate the momentum distribution of the system in the Low temperature Luttinger liquid regime $0<T\ll E_{{\rm spin}}$. The local tunneling density of states in this regime is computed in section \ref{sec:spectrum}. In section \ref{sec:spin-incoherent}, we study the properties of the system in the spin-incoherent regime $E_{{\rm spin}}<T\ll E_c$.
We summarize in the last section \ref{sec:summary}.

\section{The $SU(N)$ Hubbard Model}
\label{sec:model}

  \subsection{Hamiltonian}
  \label{subsec:Hamiltonian}

  We start with a brief review of the one-dimensional Fermionic Hubbard model  with $N$ species of spin components:~\cite{Assaraf1999}
  \begin{equation}\label{eq:H-origin}
   \mathcal{H}    =-t \sum_{i=1}^{L} \sum_{a=1}^{N} (c_{ia}^{\dag} c_{i+1a} +H.c.)    +\frac{U}{2}\sum_{i=1}^{L} \left( \sum_{a=1}^{N}n_{ia} \right)^{2}，
  \end{equation}
  where $c_{ia}$ is the Fermionic operator satisfying the commutation relation $\{c_{ia} ,c_{jb}^{\dag}\}    =\delta_{ab} \delta_{ij}$, $i$ denotes the lattice site and the spin index $a=1,\dots,N$.  The density operator $n_{ia}\equiv c_{ia}^{\dag}c_{ia}$. The first term is the non-interacting hopping term and the second term is the on-site Hubbard interaction.

   The Hamiltonian is invariant under the $SU(N)$ transformation of spin components:
   $c_{ia} \to U_{ab}c_{ib}$,
  where $U$ is any unitary $N\times N$ matrix  satisfying $U^\dag U=1$. Other than this $SU(N)$ symmetry, there is an $U(1)$ symmetry $c_{ia}\to e^{i\theta} c_{ia} $, where $\theta$ is an arbitrary constant~\cite{Assaraf1999}. This $U(1)$ symmetry results in the conservation of the particle numbers of each spin species and each species has a chemical potential $\mu_a$. In this work, we only consider the balanced case for which the chemical potential is the same for the $N$ species~\cite{Guan2012} and the number of particles for each species is the same.

The $SU(N)$ symmetric Hubbard interaction can be written as an generalized $SU(N)$ spin-spin interaction as~\cite{Assaraf1999} 
  \begin{eqnarray}
  \frac{U}{2}\left( \sum_{a=1}^{N}n_{ia} \right)^{2} =-\frac{UN}{N+1} \mathcal{S}_{i}^{A} \mathcal{S}_{i}^{A}    +\frac{U}{2} \frac{N}{N+1} \sum_{a=1}^{N}n_{ia},\nonumber\\
  \end{eqnarray}
  with 
  \begin{equation}\label{Spin}
  \mathcal{S}_{i}^{A}    =c_{ia}^{\dag} \mathcal{T}_{ab}^{A} c_{ib}, \ \ A=1,2, ...,N^2-1
  \end{equation}
being the $N^2-1$ generators of the $SU(N)$ group. The $N^2-1$ matrices $\mathcal{T}^{A}$ satisfy the commutation relation:
  \begin{equation}\label{Tcom}
    [\mathcal{T}^{A},\mathcal{T}^{B}]    =if^{ABC} \mathcal{T}^{C},
  \end{equation}
 where $f^{ABC}$ is the structure constants of the $SU(N)$ Lie algebra.

For the one-dimensional system, the Fermi surface reduces to two Fermi points and the low energy physics is determined by the processes near the two Fermi points. One can then decompose the Fermion operators  $c_{ia}$ to the left and right moving modes near the two Fermi points in the continuum limit as

  \begin{equation}\label{c_ia}
  \frac{c_{ia}}{\sqrt{\alpha}}    \rightarrow    \psi_{a}(x)    \sim    \psi_{aR}e^{ik_{F}x}  +\psi_{aL}e^{-ik_{F}x},
  \end{equation}
 where $\alpha$ is a short distance cutoff.
  
 The Hubbard interaction includes forward scatterings, back scatterings and Umklapp processes upon this decomposition. At commensurate fillings $\nu=p/q, p, q$ coprime, the Umklapp processes open a charge gap in the system and result in a Mott insulator~\cite{Assaraf1999}. However, at incommensurate filling, the Umklapp processes are absent due to momentum mismatch and the backscattering processes are also irrelevant. In this case, the system remains in the metallic phase and only the forward scatterings of the interaction are important~\cite{Capponi2016}. We then focus on this metallic phase in this work.

It was shown~\cite{Assaraf1999} that the Hamiltonian density of the Hubbard Hamiltonian Eq.(\ref{eq:H-origin}) with only forward scattering reduces to a charge part and a spin part as
 \begin{eqnarray}\label{eq:forward_hamiltonian}
      \mathcal{H}  &  =&\mathcal{H}_{c}    +\mathcal{H}_{s},\\
      \mathcal{H}_{c}    &= &\frac{\pi v_{c}}{N}  ( :\rho_{0R} \rho_{0R}:   +: \rho_{0L} \rho_{0L}:)    +G_{c} :\rho_{0R} \rho_{0L}:,\\
      \mathcal{H}_{s}    &=& \frac{2\pi v_{s}}{N+1}  ( : \mathcal{S}_{0R}^{A} \mathcal{S}_{0R}^{A}:   +: \mathcal{S}_{0L}^{A} \mathcal{S}_{0L}^{A}:)    +G_{s} :\mathcal{S}_{0R}^{A} \mathcal{S}_{0L}^{A}:.\nonumber\\
    \end{eqnarray}
    where $:...:$ denotes normal ordering and
    \begin{eqnarray}
     \rho_{0R(L)}   & =&\sum_{a=1}^{N} \rho_{aR(L)}=:\sum^N_{a=1} \psi_{aR(L)}^{\dag} \psi_{aR(L)}: ,\\
      \mathcal{S}_{0R(L)}^{A}    &=&:\psi_{aR(L)}^{\dag} \mathcal{T}_{ab}^{A} \psi_{bR(L)}:.
    \end{eqnarray}
The renormalized charge and spin mode velocities and couplings are 
    \begin{align*}
      &v_{c}    =v_{F}  +(N-1)\frac{U\alpha}{2\pi}, \ G_{c}    =-\frac{N}{N-1}U\alpha, \\
      &v_{s}    =v_{F}  -\frac{U\alpha}{2\pi}, \ G_{s}    =-2U\alpha.
    \end{align*}

For repulsive interaction $U>0$, the $G_s$ term is marginally irrelevant, we then ignore this term and study the case $G^*_s=0$ in this paper.

  \subsection{Bosonization}
  \label{subsec:GF}
The metallic phase of the one-dimensional system at low  temperature can be described by the Luttinger liquid theory. The left and right mode $\psi_{ar}$ can be Bosonized by the standard procedure:
     \begin{equation}\label{Bose-dual}
   \psi_{ar}    =\frac{U_{ar}}{\sqrt{2\pi \alpha}}e^{i(\theta_{a}-r\phi_{a})},
   \end{equation}
   where $a=1, ..., N$, $r=L,R$,  and $U_{ar}$ are Klein factors. $\theta_{a}$ and $\phi_{a}$ are two Bosonic fields satisfying the commutation relation:
   \begin{align}\label{eq:commutation}
   \begin{split}
   & [\phi_{a}(x),\phi_{b}(y)]  =[\theta_{a}(x),\theta_{b}(y)] =0,\\
   & [\phi_{a}(x),\partial_{y} \theta_{b}(y)]=i\delta_{ab} \delta(x-y).
   \end{split}
   \end{align}

   For a specific spin flavor $a$, 
    $:\psi_{ar}^{\dag}\psi_{ar}:(x)    =-\frac{1}{2\pi} (\partial_{x}\phi_{a}  -r\partial_{x}\theta_{a})$ and the charge density $\rho_a(x)=:\sum_r \psi_{ar}^{\dag}\psi_{ar}:  =-\frac{1}{\pi} \partial_{x}\phi_{a}$. The Hamiltonian Eq.(\ref{eq:forward_hamiltonian}) including the forward scatterings is quadratic in the Bosonic fields $\theta_a$ and $\phi_a$. However, the different $a$ modes are coupled. To get the normal modes of the Hamiltonian Eq.(\ref{eq:forward_hamiltonian}), one can recombine the $N$ species of bosonic modes $\theta_a$ and $\phi_a$ to a charge mode and $N-1$ spin modes as ~\cite{Assaraf1999}
 \begin{eqnarray}\label{eq:normal_mode}
     \Phi_{c} &=&   \frac{1}{\sqrt{N}} (\phi_{1}  +\phi_{2}  +\cdots   +\phi_{N}),\\
     \Theta_{c} &=&   \frac{1}{\sqrt{N}} (\theta_{1}  +\theta_{2}  +\cdots   +\theta_{N}),\\
     \Phi_{ms} &=&   \frac{1}{\sqrt{m(m+1)}} (\phi_{1}  +\cdots   +\phi_{m}  -m\phi_{m+1}),\\
     \Theta_{ms}&=&   \frac{1}{\sqrt{m(m+1)}} (\theta_{1}  +\cdots   +\theta_{m}  -m\theta_{m+1}).
\end{eqnarray}
where $m=1,2,...,N-1$.  The $N-1$ spin modes correspond to the trace of the $N-1$ Cartan generators of the $SU(N)$ group~\cite{Georgi}.
The above defined $\Phi_c, \Phi_{ms}$ and $\Theta_c, \Theta_{ms}$ satisfy the similar commutation relation Eq.(\ref{eq:commutation}).

The Hamiltonian density Eq.(\ref{eq:forward_hamiltonian}) is  diagonalized in terms of the above defined charge and spin modes as:
\begin{eqnarray}\label{eq:diagonalized-Hamiltonian}
\mathcal{H} &=& \mathcal{H}^*_c+\mathcal{H}^*_s,\\
    \mathcal{H}^*_{c} & =& \frac{u_{c}}{2\pi}   \left[\frac{1}{K_{c}}(\partial_{x}\Phi_{c})^{2}   +K_{c}(\partial_{x}\Theta_{c})^{2}\right],\\
     \mathcal{H}_{s}^{*}  &=& \frac{u_{s}}{2\pi}   \sum_{m=1}^{N-1}[(\partial_{x}\Phi_{ms})^{2}  +(\partial_{x}\Theta_{ms})^{2}],
  \end{eqnarray}
   where the Luttinger liquid parameter $K_c$ and the renormalized velocity of the charge mode in the continuum limit is respectively
   \begin{eqnarray}\label{eq:Luttinger_parameters}
      K_{c}   & =& \frac{1}{\sqrt{1+(N-1){U\alpha}/{\pi v_{F}}}},\\
      u_{c}    &=& v_{F}  \sqrt{1+(N-1){U\alpha}/{\pi v_{F}}}
   \end{eqnarray}
and $u_{s}=v_F$ at $G^*_s=0$. The corresponding Luttinger liquid parameter for the spin modes is $K_s=1$ due to the $SU(N)$ symmetry of the interaction.

   \subsection{Green's Function}
   With the diagonaized Hamiltonian Eq.(\ref{eq:diagonalized-Hamiltonian}), one can obtain the single particle Green's function in a straightforward way. The single particle Green's function is defined as 
        \begin{eqnarray}\label{eqn:single-G}
     G_{ab}(x,\tau)  &=&  -\langle \hat{T}[\psi_{a}(x,\tau) \psi_{b}^{\dag}(0,0)]\rangle\nonumber\\
     &=& e^{i k_F x}\langle e^{-i[\theta_a (x, \tau)-\theta_b(0,0)-\phi_a(x, \tau)+\phi_b(0,0)]}  \rangle\nonumber\\
     &&+e^{-i k_F x}\langle e^{-i[\theta_a (x, \tau)-\theta_b(0,0)+\phi_a(x, \tau)-\phi_b(0,0)]}  \rangle, \nonumber\\
   \end{eqnarray}
where $\tau$ is the imaginary time and $\hat{T}$ labels time-ordering.

Since the interaction  mixes the left and right mode $\psi_{aL}$ and $\psi_{aR}$, the mode $\theta_a-r\phi_a$ corresponding to $\psi_{ar}$ is not the physical left and right mode.  
However, we can obtain the physical left and right moving modes for the charge and spin modes of the diagonalized Hamiltonian Eq.(\ref{eq:diagonalized-Hamiltonian}) as
\begin{eqnarray}
\tilde{\Phi}_{cR}&=&\Phi_{c}-K_c\theta_c, \\
\tilde{\Phi}_{cL}&=&\Phi_{c}+K_c\theta_c,\\
\tilde{\Phi}_{msR}&=&\Phi_{ms}-K_s\theta_{ms}, \\
\tilde{\Phi}_{msL}&=&\Phi_{ms}+K_s\theta_{ms}.
\end{eqnarray}
The above defined modes $\tilde{\Phi}_{R(L)}$ are physically right (left) moving modes satisfying the equation of motion $\frac{\partial \tilde{\rho}_{R,L}}{\partial t}=\mp u\partial_x \tilde{\rho}_{R,L}$, where $\tilde{\rho}_{R,L}=\pm \frac{1}{2\pi}\partial_x \tilde{\Phi}_{R,L}$ and $u$ is the velocity of the corresponding charge or spin mode. The Green's functions of these modes  contain factors of only $x+ iu\tau$ or $x- iu\tau$ as~\cite{Giamarchi2004}

\begin{eqnarray}\label{eq:left_right_GF}
\langle e^{i(\tilde{\Phi}_{cr}(x,\tau)-\tilde{\Phi}_{cr}(0,0))}\rangle&=&\left(\frac{\frac{\pi\alpha}{\beta u_c}}{\sinh\frac{\pi}{\beta}(\frac{x}{u_c}-i r \tau)}\right)^{K_c},\nonumber\\
\langle e^{i(\tilde{\Phi}_{msr}(x,\tau)-\tilde{\Phi}_{msr}(0,0))}\rangle&=&\left(\frac{\frac{\pi\alpha}{\beta u_s}}{\sinh\frac{\pi}{\beta}(\frac{x}{u_s}-i r \tau)}\right)^{K_s},\nonumber\\
\end{eqnarray}
where $r=1$ for right moving mode and $-1$ for left moving mode. The correlation between different $\tilde{\Phi}$ modes and $\tilde{\Theta}$ modes vanishes since these modes decouple. 

To obtain the Green's function of Eq.(\ref{eqn:single-G}), we decompose the $\phi_a$  mode to the physically left and right moving modes $\tilde{\Phi}_{R,L}$. 
The inverse relationship between $\phi_a$ and $\Phi_c, \Phi_{ms}$ is 
\begin{eqnarray}\label{eq:normal_mode_inverse}
&&\phi_1(x, \tau)=
\frac{1}{\sqrt{N}}\Phi_{c}
+\sum_{m=1}^{N-1}\frac{1}{\sqrt{m(m-1)}}\Phi_{ms},\nonumber\\
&&\phi_a(x, \tau)=\frac{1}{\sqrt{N}}\Phi_{c}
 -\sqrt{\frac{a-1}{a}}\Phi_{a-1 s }\nonumber\\
&&\ \ \ \ \ \ \ \ \ \ \ \ \  \ +\sum_{m=a}^{N-1}\frac{1}{\sqrt{m(m-1)}}\Phi_{ms},\nonumber\\
&&\ \ \ \ \ \ \ \  \ \ \ \ \ \ \ \  (a=2, ...N-1)\nonumber\\
&&\phi_N(x, \tau)=\frac{1}{\sqrt{N}}\Phi_{c}
 -\sqrt{\frac{N-1}{N}}\Phi_{N-1 s},
\end{eqnarray}
and the same relationship holds between $\theta_a$ and $\Theta_c, \Theta_{ms}$. 
The relationship between $\phi_a, \theta_a$ and the physically left and right moving modes can be obtained by 
replacing $\Phi_c, \Phi_{ms}$ and $\Theta_c, \Theta_{ms}$ in Eqs. (\ref{eq:normal_mode_inverse}) by 
\begin{eqnarray}\label{eq:physical_mode_inverse}
\Phi_c&=&(\tilde{\Phi}_{cR}+\tilde{\Phi}_{cL})/2, \nonumber\\
\Phi_{ms}&=&(\tilde{\Phi}_{msR}+\tilde{\Phi}_{msL})/2,\nonumber\\
\Theta_c&=&(\tilde{\Phi}_{cL}-\tilde{\Phi}_{cR})/2K_c,\nonumber\\
\Theta_{ms}&=& (\tilde{\Phi}_{msL}-\tilde{\Phi}_{msR})/2K_s.
\end{eqnarray}

 Since $\langle e^{i(\sum_j A_j \tilde{\Phi}(r_j)}\rangle$ is non-vanishing only when $\sum_j A_j=0$,~\cite{Giamarchi2004}  the Green's function $G_{ab}(x,\tau)$ is non-vanishing only when $a=b$, i.e., the single-particle correlation  is zero between different spin species. From Eq.(\ref{eqn:single-G}), Eq.(\ref{eq:left_right_GF}), Eq.(\ref{eq:normal_mode_inverse}) and Eq.(\ref{eq:physical_mode_inverse}), we obtain the single-particle Green's function in the Luttinger liquid regime $0< T\ll E_{{\rm spin}}$ as 
\begin{eqnarray}\label{eq:GF-finiteT}
 &&G_{ab}(x,\tau) =\delta_{ab}\nonumber\\
       &&  \left[\frac{(\frac{\pi\alpha}{\beta u_{c}})^2}{\sinh \frac{\pi}{\beta}(\frac{x}{u_{c}}+i\tau)   \sinh \frac{\pi}{\beta}(\frac{x}{u_{c}}-i\tau)}\right]^{\frac{(1-K_c)^2}{4NK_c}} \times\nonumber\\
        &&  \left[\left(\frac{\frac{\pi\alpha}{\beta u_{c}}}   {\sinh\frac{\pi}{\beta}(\frac{x}{u_{c}}+i\tau)}\right)^{\frac{1}{N}}\left(\frac{\frac{\pi\alpha}{\beta u_{s}}}{\sinh \frac{\pi}{\beta}(\frac{x}{u_{s}}+i\tau)}\right)^{1-\frac{1}{N}} e^{-ik_{F}x} \right.\nonumber\\
       && \left.+ \left(\frac{\frac{\pi\alpha}{\beta u_{c}}}   {\sinh\frac{\pi}{\beta}(\frac{x}{u_{c}}-i\tau)}\right)^{\frac{1}{N}}\left(\frac{\frac{\pi\alpha}{\beta u_{s}}}{\sinh \frac{\pi}{\beta}(\frac{x}{u_{s}}-i\tau)}\right)^{1-\frac{1}{N}} e^{ik_{F}x}\right].\nonumber\\
\end{eqnarray}

At $T\to 0$, the above correlation function reduces to the one obtained in Ref~\cite{Assaraf1999} as
\begin{eqnarray}\label{eq:single_GF}
G_{ab}(x,\tau) &=&\delta_{ab}\left( \frac{\alpha^2}{x^2+u_c^2\tau^2} \right)^{\frac{(1-K_c)^2}{4NK_c}}\nonumber\\
&&\left[\left( \frac{\alpha}{x+i u_c\tau}\right)^{\frac{1}{N}}\left( \frac{\alpha}{x+i u_s\tau}\right)^{1-\frac{1}{N}}e^{-ik_F x}\right. \nonumber\\
&& \left. +  \left( \frac{\alpha}{x-i u_c\tau}\right)^{\frac{1}{N}}\left( \frac{\alpha}{x-i u_s\tau}\right)^{1-\frac{1}{N}}e^{ik_F x} \right].\nonumber\\
\end{eqnarray}
 The single-particle Green's function is the same for different species $a$ due to the $SU(N)$ spin symmetry.

\section{Luttinger liquid parameters}\label{sec:LL_parameters}
The Luttinger parameter $K_c$ plays an essential role in the physical properties of the system. For weak interaction $U\ll t$, the Luttinger liquid  description  in the continuum limit is valid and 
the Luttinger parameter $K_c$ is approximately given in Eq.(\ref{eq:Luttinger_parameters}). For strong interaction $U\geq t$, the Luttinger liquid description of the low energy physics of the system is not strictly valid. The accepted view is that the effect of strong interaction still corresponds to a renormalization of the Luttinger parameters $K_c$ and $u_c$ as well as the spin velocity $u_s$~\cite{Assaraf1999, Kawakami1993}, which however is different from Eq.(\ref{eq:Luttinger_parameters}).  These parameters completely characterize the low energy properties of the metallic phase and can be obtained from the Bethe Ansatz solution in the strong interaction case. Though  the one-dimensional SU(N) Hubbard model on a lattice is not exactly solvable by Bethe-Ansatz for $N>2$ or when scattering processes with three or more Fermions on one site take place~\cite{Assaraf1999, Frahm1993}, in the metallic phase with small densities or large interaction $U$,  the three or more particle scattering processes are negligible and the Bethe Ansatz equations describe the system reasonable well. Actually,  in the latter case, the metallic phase of the 1D $SU(N)$ Hubbard model reduces to a 1D Fermionic gas with $\delta$ interaction which again is solvable by Bethe-Ansatz~\cite{Sutherland1968, Schlottmann1993, Guan2012}. 

For this purpose, the Luttinger parameter $K_c$ can be related to the compressibility of the system by the relationship~\cite{Giamarchi2004}
\begin{equation}
\kappa=-\frac{1}{L}\frac{\partial L}{\partial P}=\frac{1}{\bar{\rho}_c^2}\frac{\partial \langle\rho(x)\rangle}{\partial \mu}=\frac{1}{\bar{\rho}_c^2}\frac{NK_c}{\pi u_c},
\end{equation}
i.e., 
\begin{equation}\label{eq:compressibility}
 \kappa^{-1}=\frac{1}{L}\frac{\partial^2 E}{\partial \langle\rho(x)\rangle^2}=\bar{\rho}_c^2\frac{\pi u_c}{N K_c}, 
\end{equation}
where $\rho(x)=\bar{\rho}_c+ \rho_0(x)=\bar{\rho}_c-\frac{1}{\pi}\sum_a \partial_x \phi_a$ is the local number density of the system, $\bar{\rho}_c= \langle\rho(x)\rangle=N k_F/\pi$ is the average number density of the whole chain, 
and $E$ is the ground state energy per unit length which can be obtained from the Bethe Ansatz solution~\cite{Guan2012}.

The velocity of the charge mode $u_c$ can be expressed as~\cite{Haldane1981, Frahm1990, Frahm1993}
\begin{equation}\label{eq:velocity}
u_c=\frac{1}{2\pi n_c(k_F)}\epsilon'_c(k_F).
\end{equation}
Here $n_c(k)$ is the  charge/particle distribution function with momentum $k$, and $\epsilon_c(k)$ is the dressed energy of the interacting system defining the energy bands of the system~\cite{Haldane1981, Frahm1990, Frahm1993}. Both $n_c(k)$ and $\epsilon_c(k)$ satisfies a set of coupled Bethe Ansatz integral equations and can be solved numerically~\cite{Frahm1993}. Combining Eq.(\ref{eq:compressibility}) and (\ref{eq:velocity}), the Luttinger parameter $K_c$ can then be obtained  numerically from the Bethe Ansatz solution.

It was also discovered from the Bethe Ansatz equations that at zero magnetic field the compressibility is related to the dressed charge $z_c$ as~\cite{Frahm1990}
\begin{equation}
z^2_c=\pi u_c \bar{\rho}_c^2 \kappa=NK_c,
\end{equation}
where the dressed charge $z_c$ characterizes the number of electrons added near the Fermi surface due to a change in the total density $\bar{\rho}_c$ and is defined as~\cite{Frahm1990}
\begin{equation}
z_c\equiv \xi_c(z_0)=\frac{1}{2n_c(k_F)}\frac{\partial \bar{\rho}_c}{\partial k_F}
\end{equation}
 with $z_0\equiv \sin k_F/\tilde{U}$ and $\tilde{U}\equiv U/t$. The function $\xi_c(z)$ satisfies the following integral equation at zero magnetic field~\cite{Frahm1993}
\begin{equation}\label{eq:dressed-charge}
\xi_c(z)=1+\frac{1}{2\pi}\int^{z_0}_{-z_0}dy \xi_c(y)G_N(z-y;1)
\end{equation}
with 
\begin{equation}
G_N(x_1;x_2)=\frac{1}{Nx_2}{\rm Re}[\psi(1+i\frac{x_1}{2Nx_2})-\psi(\frac{1}{N}+i\frac{x_1}{2Nx_2})],
\end{equation}
where $\psi(x)$ is the digamma function.

 \begin{figure}[tpb]
\includegraphics[width=7.8cm]{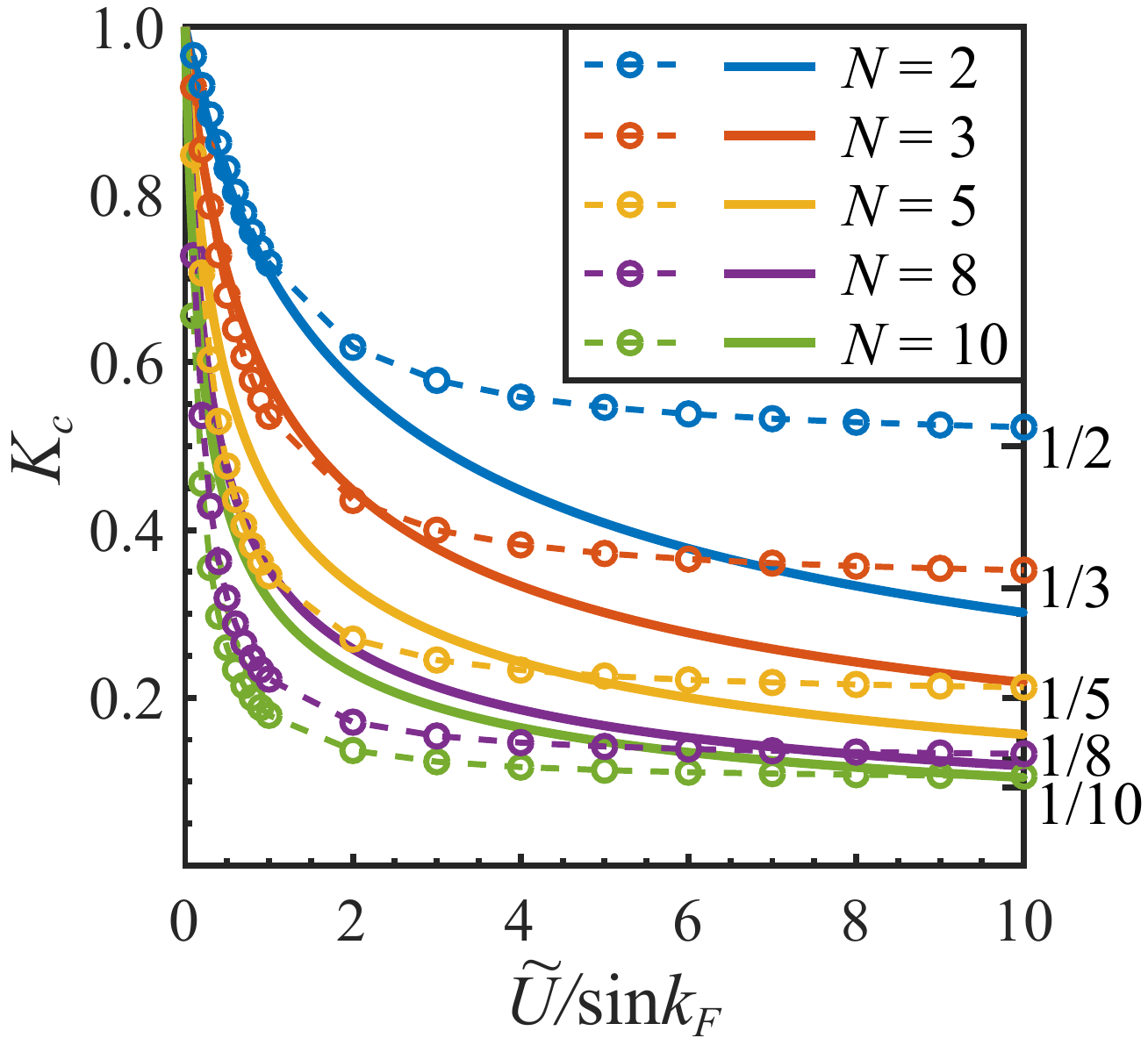}
\caption{The Luttinger liquid parameter $K_c$ as a function of interaction $\tilde{U}$ and charge density (which determines $k_F$) for different $N$. The solid lines are plots from Eq.(\ref{eq:Luttinger_parameters}) and the dashed lines with circles are results from Bethe-Ansatz Eq.(\ref{eq:dressed-charge}).
}\label{fig:kc}
\end{figure}

The above Eq.(\ref{eq:dressed-charge}) can be solved numerically for arbitrary $z_0$, i.e., arbitrary interaction $U$ and charge density $\bar{\rho}_c$, from which we can then get the Luttinger parameter  $K_c$ for arbitrary interaction and particle density. Figure~\ref{fig:kc} shows the Luttinger liquid parameter $K_c$ as a function of $1/z_0=\tilde{U}/{\rm sin} k_F$ for different $N$ obtained by solving Eq.(\ref{eq:dressed-charge}) numerically. One can see that at weak interactions, the results from the Bethe-Ansatz agree with the ones from Luttinger liquid theory Eq.(\ref{eq:Luttinger_parameters}), whereas at strong interaction, the results from the Bethe-Ansatz deviate obviously from Eq.(\ref{eq:Luttinger_parameters}).

At small $z_0\ll 1$, i.e., strong interaction $U$,  the solution of $z_c$ can be obtained by iteration of Eq.(\ref{eq:dressed-charge}) as 
\begin{equation}\label{eq:zc_largeU}
z_c\approx 1+ z_0 [G_N(0;1)]/\pi=1+z_0[\gamma +\psi(1/N)]/N\pi,
\end{equation}
where $\gamma$ is the Euler constant. We then see that at $U\to \infty$, $z_0\to 0$, $z_c\to 1$ and the Luttinger parameter $K_c\to 1/N$ as shown in Fig.\ref{fig:kc}.

At large $z_0\gg N$, a perturbative scheme based on the Wiener-Hopf method can be applied to Eq.(\ref{eq:dressed-charge}) and~\cite{Frahm1993} 
\begin{equation}\label{eq:zc_smallU}
z_c\approx \sqrt{N} (1-\frac{N-1}{2\pi z_0}).
\end{equation}
We see that at $U\to 0$, $z_0\to \infty$, $z_c\to \sqrt{N}$, the Luttinger parameter $K_c$ then reduces to the non-interacting case $K_c=1$ as also shown in Fig.\ref{fig:kc}.

\section{Momentum distribution}\label{sec:momemtum_distribution}
One of the measurable observables which reveals the correlation of the system is the momentum distribution $n_c(k)$ of particles. It can be measured in cold atom experiments by extinguishing the trapping light and imaging the atomic cloud after a ballistic expansion~\cite{Fallani2014}. In a recent experiment on the one dimensional  $Yb^{173}$ atoms, the momentum distribution of a one-dimensional system with $SU(N)$ spin symmetry was measured  for different $N$~\cite{Fallani2014}. In this section, we compute the momentum distribution of the $SU(N)$ Luttinger liquid at low temperature, i.e., the Luttinger liquid regime $0< T\ll E_{{\rm spin}}$, and compare the results with the experiments~\cite{Fallani2014}. The spin-incoherent case at intermediate temperature $E_{{\rm spin}}\ll T\ll E_c$ will be discussed in a later section.

The momentum distribution 
  can be expressed in terms of the time-ordered Green's function as
  \begin{equation}\label{def-Momentum-dis}
    n_c(k)  =\sum_a \int_{-\infty}^{\infty}dx G_{aa}(x,\tau \rightarrow 0^{-}) e^{-ikx}.
  \end{equation}
  
 In the Luttinger liquid regime $0< T\ll E_{{\rm spin}}$, the temperature  dependence of the momentum distribution is weak and we can take the limit $T\to 0$. The momentum distribution is then obtained as 
 \begin{equation}
 n_c(k)=n_{R} (k) +n_{L}(k)
 \end{equation}
where the left and right branches satisfies  $n_L(k)=n_R(-k)$. We then only need to calculate the right branch:
\begin{eqnarray}\label{eq:momentum_distribution}
n_{R}(k)&=&N\alpha^{\tilde{K}} (2\alpha)^{-\frac{\tilde{K}}{2} -\frac{1}{2}}  | k -k_{F}|^{\frac{\tilde{K}}{2} -\frac{1}{2}} \nonumber  \\
              && \times  \{ \theta(k -k_{F}) \frac{1}{\Gamma(\frac{\tilde{K}}{2})}  W_{-\frac{1}{2},-\frac{\tilde{K}}{2}}[2\alpha(k -k_{F})]\nonumber\\
              && +\theta(-k +k_{F})  \frac{1}{\Gamma(\frac{\tilde{K}}{2} +1)}  W_{\frac{1}{2},-\frac{\tilde{K}}{2}}[-2\alpha(k -k_{F})]\},\nonumber\\
 \end{eqnarray}
 where
  $\tilde{K} \equiv \frac{1}{2NK_{c}}(K_{c}-1)^{2} $
  varies from $0$ to $1/2$ for the interaction $U$ changing from $0$ to $\infty$,
 and   $W_{\lambda,\mu}(z)$ is Whittaker function:
  \begin{align}\label{W}
  \begin{split}
    W_{\lambda,\mu}(z)  =&\frac{\Gamma(-2\mu)}{\Gamma(\frac{1}{2} -\mu -\lambda)}M_{\lambda,\mu}(z)  \\
                         &\quad +\frac{\Gamma(2\mu)}{\Gamma(\frac{1}{2} +\mu -\lambda)}M_{\lambda,-\mu}(z).
  \end{split}
  \end{align}
  $M_{\lambda,\mu}(z)$ can be written in terms of the confluent Hypergeometric function $\Phi(a,b,z)$:
  \begin{eqnarray}
     M_{\lambda,\mu}(z)  & =& z^{\mu +\frac{1}{2}}  e^{-\frac{z}{2}}  \Phi(\mu -\lambda +\frac{1}{2},\ 2\mu +1,\ z),\nonumber\\
     M_{\lambda,-\mu}(z)  &  = & z^{-\mu +\frac{1}{2}}  e^{-\frac{z}{2}}  \Phi(-\mu -\lambda +\frac{1}{2},\ -2\mu +1,\ z).\nonumber\\
 \end{eqnarray}

\begin{figure}[tpb]
\includegraphics[width=7.5cm]{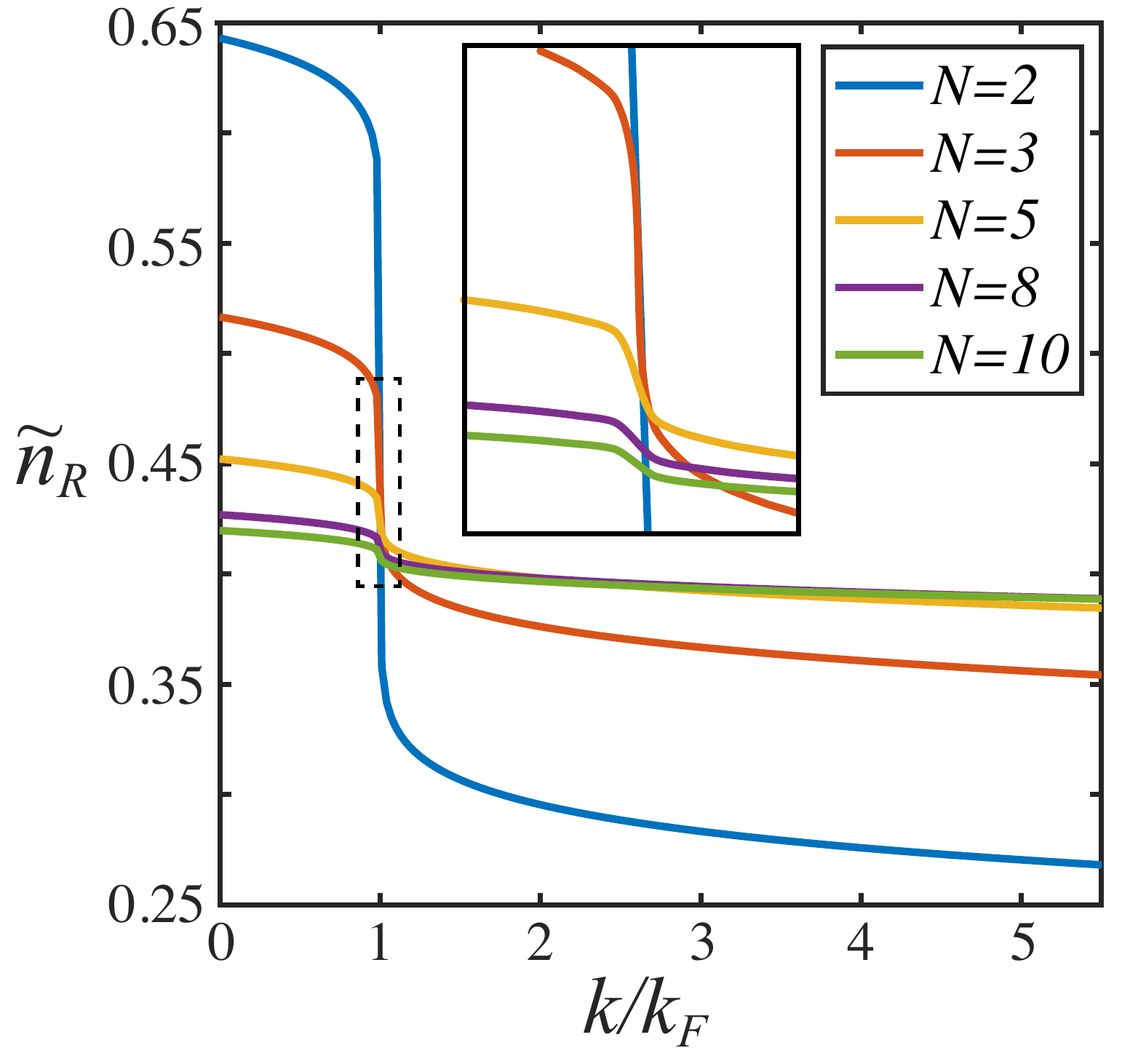}
\caption{The momentum distribution per spin species $\tilde{n}_R(k) \equiv n_R(k)/N$ for a typical case with $\tilde{U}=5, \sin k_F=0.9$.
The inset shows the zoomed-in picture of the dashed area near the Fermi momentum $k_F$.
}\label{fig:momen}
\end{figure}

 The momentum distribution per spin species $\tilde{n}_R(k) \equiv n_R(k)/N$ for a typical case with $\tilde{U}=5, \sin k_F=0.9$ from Eq.(\ref{eq:momentum_distribution})
 is shown in Fig. \ref{fig:momen}.

 Expanding the expression in Eq.(\ref{eq:momentum_distribution}) near the Fermi points, we get near $k=k_F$, the momentum distribution of the right branch is 
 \begin{eqnarray}\label{eq:momentum_distribution_LL}
    n_{R}(k) &&= \frac{ N}{\tilde{K}}  \frac{ 1}{ B(\frac{1}{2},  \frac{\tilde{K}}{2})}\nonumber\\
           && +\frac{ N \alpha^{\tilde{K}}}{\tilde{K}} \frac{\Gamma(1-\tilde{K})\sin( \frac{\pi \tilde{K}}{2})}{\pi}  \Sign(k -k_{F})  |k -k_{F}|^{\tilde{K}}.\nonumber\\
   \end{eqnarray}

In the non-interacting case, $K_c=1, \tilde{K}=0$, the momentum distribution $n_R(k)$ reduces to 
\begin{equation}
n_R(k)=\frac{N}{2}[1-\Sign(k -k_{F})],
\end{equation}
which shows a sharp Fermi surface at $k=k_F$. 

In the finite interaction case, we obtain $0<\tilde{K}\leq 1/2$ for the interaction varying from $0$ to $\infty$ from the analysis of the last section. The momentum distribution in Eq.(\ref{eq:momentum_distribution_LL}) then becomes continuous at $k=k_F$ and the Fermi surface disappears at $T=0$ as shown in Fig.\ref{fig:momen}. This indicates the breakdown of single particle excitations in one dimensional interacting systems and all the excitations are collective in such systems.

The momentum distribution has a power law singularity as constant$+|k-k_F|^{\tilde{K}}\Sign(k -k_{F})$ at $k=k_F$. 
The Luttinger liquid parameter $K_c$ decreases with increasing $N$ at fixed interaction and density as shown in Fig.\ref{fig:kc} so the power index $\tilde{K}\equiv \frac{1}{2NK_{c}}(K_{c}-1)^{2}$  increases with $N$ at fixed interaction and  density. 
The momentum distribution $n_R(k)$ then becomes more and more broadened with the increase of $N$ for fixed interaction as shown in Fig.\ref{fig:momen}. 
This  is consistent with the experimental observations in the one-dimensional alkaline-earth atomic systems in Ref.~\cite{Fallani2014}, 
and also indicates that the one-dimensional interacting Fermionic system deviates more and more from a Fermi liquid with the increase of $N$.

Interestingly, it was pointed out  that the ground state energy per particle for an $N$-component one-dimensional Fermionic system with $N\to \infty$ is the same as for  a  one-dimensional spinless Bosonic system with the same repulsive  $\delta$-interaction and total number density~\cite{Yang2011} in the limit of large number density, indicating Bosonic behavior for the multi-component Fermionic system with $N\to \infty$, which was  observed in experiments~\cite{Fallani2014}.
However, we note here that the single-particle Green's function and so the momentum distribution for the two systems are not the same due to the different statistics they obey. This is clear by a direct comparison of Eq.(\ref{eq:single_GF}) in the  $N\to \infty$ limit
with the single-particle Green's function of a one-dimensional spinless Bosonic system with the same repulsive interaction, which is~\cite{Giamarchi2004}
\begin{equation}
\langle \psi^\dag_B(r)\psi_B(0)\rangle\sim \left(\frac{\alpha}{r}\right)^{\frac{1}{2K_B}}.
\end{equation}
Here $r=(x, u\tau)$ and $K_B$ is the Luttinger liquid parameters for the one-dimensional spinless Bosonic system. For local repulsive interaction $U$, $K_B$ is $+\infty$ at $U=0$, indicating a long range order due to Bose condensation. At finite repulsion, $K_B$ is finite and decreases with the increase of $U$, until at $U\to \infty$, $K_B$ approaches $1$~\cite{Giamarchi2004}. We then see that at fixed interaction, the single particle Green's function for the one-dimensional spinless Bosonic system and N-component Fermionic system with $N\to \infty$ is not the same. The latter always decays faster than the former with the same interaction. For example, at $U\to \infty$, the Green's function for the spinless Bosonic system behaves as $(1/r)^{\frac{1}{2}}$ whereas the Green's function for the N-component Fermionic system behaves approximately as $\sim (1/r)^{1+(1-1/N)^2/2}$, which always decays faster than the spinless Bosonic system.

\section{Spectrum function and tunneling density of states}\label{sec:spectrum}
Another important physical observable is the spectrum function $A(k,\omega)=-2{\rm Im}G^R(k, \omega)$ which is  related to the local tunneling density of states (DOS) as
\begin{equation}
\rho(\omega)=-\sum_a{\rm Im}\int \frac{dt}{\pi} \  e^{i\omega t} \ G^R_a(x=0, t)=\frac{1}{L}\sum_{k,a} A_a(k, \omega).
\end{equation}
The retarded Green's function $G_a^R(x=0, t)$ can be obtained from the time-ordered single-particle Green's function $G_a(x=0, t)$ by the relationship
\begin{eqnarray}
G_a^R(x=0, t)&=&i\theta(t)[G_a(x=0, t)-G^*_a(x=0, -t)]\nonumber\\
&=&-2\theta(t) {\rm Im}\ G_a(x=0, t)
\end{eqnarray}
where $G^*_a(x=0, -t)$ is the complex conjugate of $G_a(x=0, -t)$ and the real time Green's function can be obtained from the imaginary time Green's function by the Wick rotation $\tau=it+\epsilon\ {\rm sgn(t)}$.

In the Luttinger liquid regime, one can take the limit $T\to 0$ since $T\ll E_{{\rm spin}}\ll E_c$.
 The tunneling DOS for the $a$ species is then
   \begin{eqnarray}
    \rho_a(\omega) &=& \rho_{aL}(\omega) + \rho_{aR}(\omega),\nonumber\\
     \rho_{aL}(\omega) &=&  \frac{i \alpha^{\tilde{K}}}{2\pi } \int_{-\infty}^{\infty} dt e^{i \omega t}   \left[\frac{1}{-(u_{c}t - i\epsilon )^{2}}\right]^{\frac{1}{2} \tilde{K}} \nonumber\\
                   &&\quad \times \left(\frac{1}{- u_{c}t + i\epsilon }\right)^{\frac{1}{N}}   \left(\frac{1}{- u_{s}t + i \epsilon }\right)^{1-\frac{1}{N}},\nonumber\\
                   \rho_{aR}(\omega) &=& \rho_{aL}(\omega),
    \end{eqnarray}
    where  $\rho_{aL}(\omega)$ and  $\rho_{aR}(\omega)$ are the contributions from the left and right branch of the Green's function  respectively.
    From the following equation
    \begin{align}\label{ET I 118(4)}
    \begin{split}
    \int_{-\infty}^{\infty} (b+ ix)^{-\nu} e^{-ipx} dx =\theta(-p) \frac{2\pi (-p)^{\nu -1}e^{bp}}{\Gamma(\nu)}, &\\
    [\re \nu>0,\ \re b>0],&
    \end{split}
    \end{align}
    we then get the tunneling DOS  at low temperature $k_BT<\omega$ as
    \begin{equation}\label{DOS+02}
     \rho(\omega) =  \frac{ N\alpha^{\tilde{K}}}{\Gamma(\tilde{K}+1) }  \frac{1}{u_{c}^{\tilde{K}+ \frac{1}{N}} u_{s}^{1- \frac{1}{N}}}  |\omega|^{\tilde{K}}.
    \end{equation}
    
    The local tunneling DOS is  a power law of energy with power index $\tilde{K}=\frac{1}{2NK_c}(K_c-1)^2$, which increases with the increase of $N$ and the interaction, and is measurable in experiments, such as the tunneling conductance in Ref.~\cite{Bockrath1999} We see that in the Luttinger liquid regime, the power index  $\tilde{K}$ of the tunneling DOS is always positive in the interacting case $K_c<1$, resulting in a suppression of the tunneling DOS at low energy~\cite{Giamarchi2004, Glazman1992, Gogolin1998}, whereas in the non-interacting case, $\rho(\omega)$ remains a constant. This suppression of the DOS in the Luttinger liquid is a manifestation of the orthogonality catastrophe effect due to the interaction~\cite{Giamarchi2004, Glazman1992, Gogolin1998, Fiete2007}.
    We will see that when the temperature increases and the system enters the spin-incoherent regime $E_{{\rm spin}}\ll T\ll E_c$, this result will change significantly.

\section{Spin-incoherent regime}\label{sec:spin-incoherent}
In the strong interacting limit,  the charge mode velocity $u_c$ of the one-dimensional interacting system is much faster  than the velocity of the spin mode $u_s$. There then exists a spin-incoherent regime $v_s k_F=E_{{\rm spin}}< T <E_c= v_c k_F$ where the charge mode can still be described by the Luttinger liquid theory, however, the Luttinger liquid description of the spin degrees of freedom is no longer valid~\cite{Fiete2004, Fiete2007, Cheianov2004, Cheianov2005}.

 The correlation function in this regime does not depend on the spin part of the Hamiltonian since the spin  Hamiltonian $H_s$ drops from the partition function $Z\equiv {\rm Tr} [e^{-\beta H} ]$ at $E_{\rm spin}\ll T$~\cite{Fiete2004, Fiete2007}. The spin configuration in this regime is completely disordered due to the thermal excitations, whereas the charge part can still be described by the Luttinger liquid Hamiltonian Eq.(\ref{eq:diagonalized-Hamiltonian}). The asymptotic behavior of the single-particle Green's function in the spin-incoherent regime can be easily obtained by generalizing the $SU(2)$ case in Ref~\cite{Fiete2004, Fiete2007} to the $SU(N)$ case here.

Since the charge and spin parts are separated and the spin Hamiltonian drops from the partition function, it's convenient to define a spinless charge mode corresponding to the total charge or particle number of the system as
\begin{eqnarray}
\Phi&=&\sum_a \phi_a=\sqrt{N} \Phi_c,\\
\Theta&=&\sum_a \theta_a/N=\Theta_c/\sqrt{N},
\end{eqnarray}
which satisfy the commutation relation $[\Phi(x_1), \nabla\Theta(x_2)]=i\pi \delta(x_1-x_2)$.  The total charge density of the system is then 
\begin{equation}
\rho(x)=\bar{\rho}_c-\frac{1}{\pi}\partial_x \Phi,
\end{equation}
where $\bar{\rho}_c=N k_F/\pi$ is the average charge or particle density.
The Hamiltonian density of the charge part of Eq.(\ref{eq:diagonalized-Hamiltonian})  becomes
\begin{equation}\label{eq:charge_Hamiltonian}
    \mathcal{H}_{c}  =\frac{u_{c}}{2\pi}   \left[\frac{1}{NK_{c}}(\partial_{x}\Phi)^{2}  +NK_{c}(\partial_{x}\Theta)^{2}\right],
 \end{equation}
in terms of $\Phi$ and $\Theta$.

The single-particle Green's function $G_{a}(x,\tau)  =  -\langle \hat{T}[\psi_{a}(x,\tau) \psi_{a}^{\dag}(0,0)]\rangle$  describes the probability of a particle inserted at $(0,0)$ propagating to $(x, \tau)$ and annihilated. 
It can be expressed as a path integral over all possible path from the initial  position in space-time $(0,0)$ to the final position $(x, \tau)$ as 
\begin{eqnarray}\label{eq:GF_integral}
G_{a}(x,\tau) &=&\frac{1}{Z}{\rm Tr}[e^{-\beta H} \psi_{a}(x,\tau) \psi_{a}^{\dag}(0,0)]\nonumber\\
&=&\frac{1}{Z} \int {\cal D}\psi \ e^{- S_E}\psi_{a}(x,\tau) \psi_{a}^{\dag}(0,0),
\end{eqnarray}
where  the partition function $Z=\int {\cal D}\psi \ e^{ -S_E}$ is a path integral over the probability of all the paths.

 \begin{figure}[b]
\includegraphics[width=7cm]{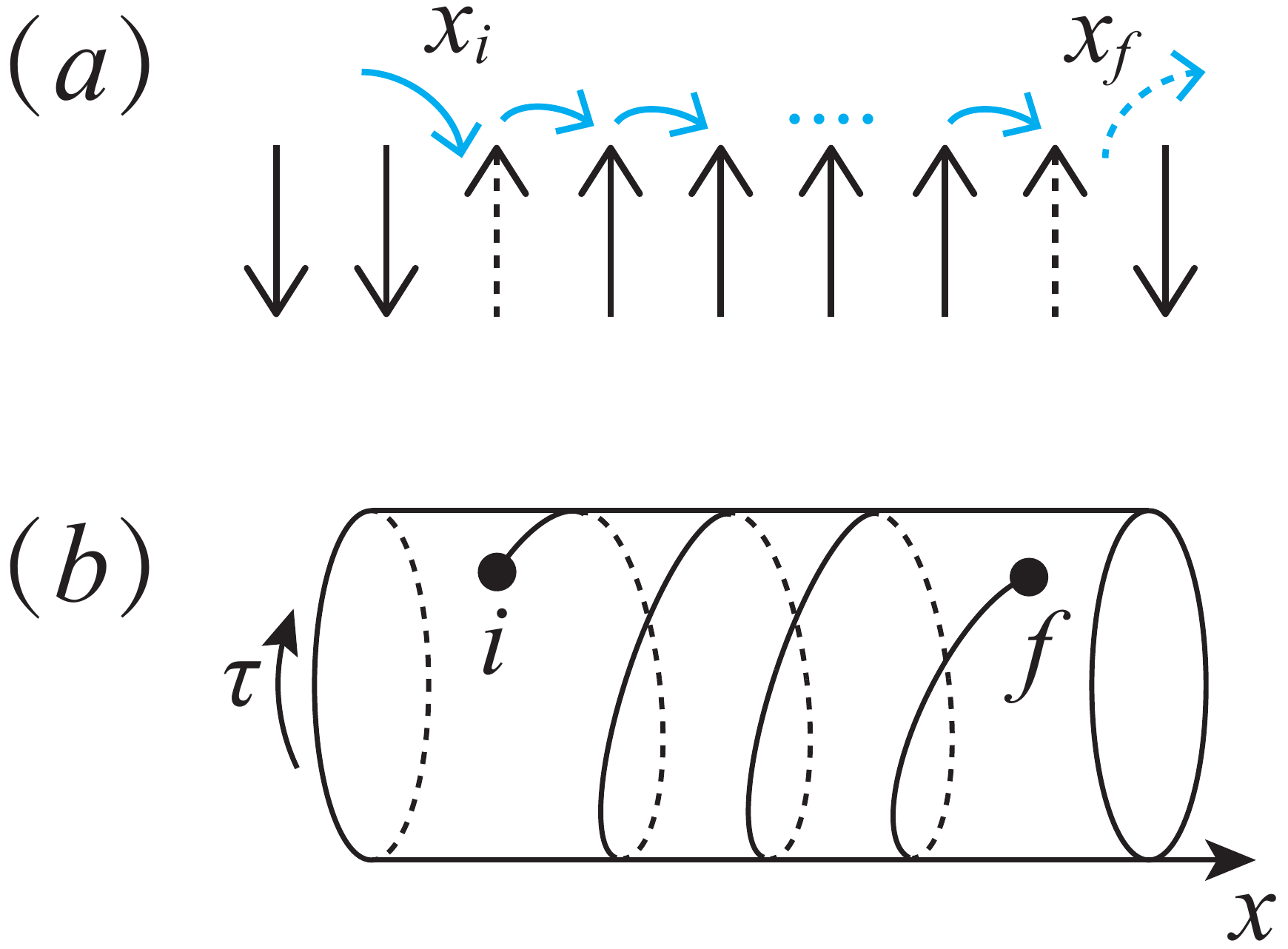}
\caption{(a)The most probable process for a Fermion propagating from the initial position to the final position in the spin-incoherent regime is through permutation with identical neighboring particles. (b)The corresponding worldline for the process in (a) wraps around the imaginary time axis from the initial to final position. The circumference of the imaginary time axis is $\beta=1/T$.
}\label{fig:worldlines}
\end{figure}

In the spin-incoherent regime, the crossing of two particles is 
 almost prohibited since the crossing time $\sim 1/E_{{\rm spin}}$ is much larger than the typical thermal coherence time $\beta=1/k_B T$~\cite{Fiete2004, Fiete2007}. For large distance $(x,\tau)$, the most probable process to insert a particle with spin species $a$ at $(0,0)$ and remove it at $(x,\tau)$ is through permutation of  identical neighboring particles one by one from $(0,0)$ to $(x, \tau)$, as shown in Fig.\ref{fig:worldlines}(a).  The path of the particle propagating from $(0,0)$ to $(x,\tau)$ is then equivalent to a continuous world-line wrapping around the periodic imaginary time axis as shown in Fig.\ref{fig:worldlines}(b) (see also Fig.2 of Ref~\cite{Fiete2007}). 
 Such a process requires that all the particles permuted with the initial one have the same spin species.
 The topology of the paths in Fig.\ref{fig:worldlines}(a) is then identical to that of spinless fermions~\cite{Fiete2004, Fiete2007}.

The weight for the path of Fig.\ref{fig:worldlines}(a) then includes the following contributions~\cite{Fiete2004, Fiete2007}: one is the Euclidean action factor $e^{-S_E}$ corresponding to the spinless charge Hamiltonian Eq.(\ref{eq:charge_Hamiltonian}) for each path; the other is a statistical factor $N^{-l(x, \tau)}$ coming from the requirement that all the particles between $(0,0)$ and $(x, \tau)$ have the same spin component, where $l(x,\tau)$ is the number of particles in between $(0,0)$ and $(x, \tau)$. Other than this, there is a  factor $(-1)^{l(x,\tau)}$ coming from the permutation of Fermions in between $(0,0)$ and $(x, \tau)$. The single-particle Green's function in the spin-incoherent regime for large $\tau$ is then approximately~\cite{Fiete2004, Fiete2007}
\begin{eqnarray}\label{eq:GF_incoherent_0}
G_a(x, \tau)&\sim& \sum_{m=-\infty}^\infty  \delta(m-l(x, \tau)) \langle N^{-l(x, \tau)}\nonumber\\
&&\ \ \ \ \ \ (-1)^{l(x,\tau)}  e^{i(\Theta(x,\tau)-\Theta(0,0))} \rangle,
\end{eqnarray}
where $e^{i(\Theta(x,\tau)-\Theta(0,0))}$ creates a particle at $(0,0)$ and annihilates it at $(x, \tau)$, and the average is done with respect to the charge Hamiltonian Eq.(\ref{eq:charge_Hamiltonian}).

The number of particles between the initial and final position is related to the spinless charge mode as
\begin{equation}
l(x,\tau)=\bar{\rho}_c x-\frac{1}{\pi}(\Phi(x, \tau)-\Phi(0,0)).
\end{equation}

In the limit of large $x$, $l(x, \tau)$ is large and one can turn the sum in the above equation to an integral over $m$ and get
\begin{eqnarray}
G_a(x, \tau)&\sim&   \langle N^{-l(x, \tau)}(-1)^{l(x,\tau)} e^{i(\Theta(x,\tau)-\Theta(0,0))} \rangle,\nonumber\\
\end{eqnarray}
where  $(-1)^{l(x,\tau)}={\rm Re}[e^{i\pi l(x, \tau)}]=\frac{e^{i\pi l(x, \tau)}+e^{-i\pi l(x, \tau)}}{2}$.

Denoting $\tilde{\Phi}=\Phi(x, \tau)-\Phi(0,0)$ and $\tilde{\Theta}=\Theta(x, \tau)-\Theta(0,0)$, and applying the correlation function
$\langle \tilde{\Phi}^2\rangle=\frac{NK_c}{2} {\rm ln}(x^2+u_c^2 \tau^2)$, $\langle \tilde{\Theta}^2\rangle=\frac{1}{2NK_c} {\rm ln}(x^2+u_c^2 \tau^2)$, and $\langle \tilde{\Phi}\tilde{\Theta}\rangle=\frac{1}{2} {\rm ln}\frac{u_c\tau-ix}{u_c\tau+ix}$ for the spinless charge mode Hamiltonian Eq.(\ref{eq:charge_Hamiltonian}), the single-particle Green's function $G_a(x, \tau)$ in the spin-incoherent regime is then
\begin{eqnarray}\label{eq:GF_incoherent}
G_a(x, \tau)&\sim&e^{-N k_F x {\rm ln} N/\pi} \nonumber\\
&& \langle (e^{iN k_F x-i(1+i{\rm ln} N/\pi)(\Phi(x, \tau)-\Phi(0,0))}+h.c.)\nonumber\\
&&\ \ \ e^{i(\Theta(x, \tau)-\Theta(0,0))}\rangle,\nonumber\\
&\sim&\frac{e^{-N k_F x \frac{\ln N}{\pi}}}{(x^2+u_c^2\tau^2)^{\Delta_c}}\nonumber\\
&&\times\left(\frac{e^{i(N k_F x-\phi^+_c)}}{u_c \tau-ix}+\frac{e^{-i(N k_F x-\phi^-_c)}}{u_c \tau+ix}\right),
\end{eqnarray}
where the power index 
\begin{eqnarray}
\Delta_c=\frac{1}{4}NK_c\left[1-\left(\frac{{\rm ln} N}{\pi}\right)^2\right]+\frac{1}{4NK_c}-\frac{1}{2},
\end{eqnarray}
and the phase
\begin{eqnarray}
\phi^\pm_c&=&\frac{\ln N}{\pi}\left[\frac{1}{2}NK_c \ln(x^2+u_c^2\tau^2)
\pm\frac{1}{2}{\rm ln}\left(\frac{u_c\tau-ix}{u_c\tau+ix}\right)\right].\nonumber\\
\end{eqnarray}
For the case $N=2$, the above results reduce to the $SU(2)$ case with $S=1/2$ in Ref~\cite{Fiete2004, Fiete2007}.

From Eq.(\ref{eq:GF_incoherent}), we see that the Green's function in the spin-incoherent regime decays exponentially with distance due to the disordered configuration of spins. 
There is also an oscillating phase factor in the correlation function coming from the propagation of the charge mode, however, the wave vector is $N k_F$ instead of $k_F$ as in the Luttinger liquid regime.
The spectrum function at low frequency in the spin-incoherent regime  then show a center at $\omega=u_c(q\pm N k_F)$ which is completely broadened with width $\sim NK_F {\rm ln} N$. In contrast, in the Luttinger liquid regime, the spectrum function shows peaks at $\omega=u_c(q\pm k_F)$ with power law singularity~\cite{Giamarchi2004}.

The  Green's function obtained in Eq.(\ref{eq:GF_incoherent}) is valid for large $x$ in the spin-incoherent regime. To obtain the tunneling DOS in this regime, one needs to calculate the Green's function at $x\to 0$.
The expression Eq.(\ref{eq:GF_incoherent_0}) is still valid  for small $x$ and large $\tau$ in the spin-incoherent regime, but the sum in Eq.(\ref{eq:GF_incoherent_0}) can no longer turn into an integral. 
Following the sum rule in Ref~\cite{Fiete2007} for small $x$, 
we get the single particle Green's function at $x\to 0$ as
\begin{eqnarray}
G_a(0, \tau)\sim \sqrt{\frac{\pi}{2\langle \tilde{\Phi}^2\rangle}}e^{-\langle \tilde{\Theta}^2\rangle/2}\sim\tau^{-\frac{1}{2NK_c}}\sqrt{\frac{\pi}{2NK_c {\rm ln} (u_c \tau)}}.\nonumber\\
\end{eqnarray}

The tunneling density of states in the spin-incoherent regime is then 
\begin{equation}\label{eq:DOS_incoherent}
\rho(\omega)\sim \omega^{\frac{1}{2NK_c}-1}/\sqrt{|{\rm ln}\ \omega|}.
\end{equation}
 For the short range Hubbard interaction, $1/N \leq K_c\leq 1 $ as presented in Section III. The tunneling DOS then diverges at $\omega \to 0$ for any interaction and any $N$ in the spin-incoherent regime. This is in contrast to the Luttinger liquid regime where the tunneling DOS goes to zero at $\omega \to 0$ as shown in the last section. The reason  is due to the large number of excited spin modes  in the spin-incoherent regime which beats the orthogonality catastrophe coming from the charge degrees of freedom~\cite{Fiete2007}.

It's interesting to note from Eq.(\ref{eq:zc_largeU}) and Eq.(\ref{eq:zc_smallU}) that the power index of $\rho(\omega)$ in Eq.(\ref{eq:DOS_incoherent}) increases with the increase of $N$ at strong interaction so the tunneling DOS is less and less singular at larger $N$, whereas it decreases with the increase of $N$ at weak interaction so the tunneling DOS is more and more singular with the increase of $N$ in this case. In contrast, in the Luttinger liquid regime, the power index $\tilde{K}$ of the tunneling density of states increases with increasing $N$ at both weak and strong interaction.

The above scheme to obtain the Green's function in the spin-incoherent regime is valid only for large $\tau\gg x/u$ in principle.
 To compute the momentum distribution in the spin-incoherent regime, one needs the Green's function at $\tau \to 0$ and arbitrary $x$, i.e., the density matrix $\rho_a(x)=\langle \psi^\dag_a(x)\psi_a(0)\rangle$ for any $x$, which is however not available from the current scheme. The momentum distribution in the spin-incoherent regime  then remains a challenge in a general case.

One special situation is when the local repulsion $U$ goes to infinity. In this case,  the exact eigenfunctions of the system from the Bethe Ansatz solution factorizes into a charge part similar to the wave function of noninteracting spinless fermions and a spin part which corresponds to the eigenstates of isotropic Heisenberg chain~\cite{Ogata1990, Cheianov2004, Cheianov2005}. In the spin-incoherent regime, the spin part is disordered and all the spin configurations have the same possibility. This greatly simplifies the density matrix, and an exact solution of which  is accessible by expressing the density matrix in terms of a Fredholm determinant~\cite{Cheianov2005}. Cheianov et al obtained the Green's function and density matrix in the spin-incoherent regime at  infinite repulsion  for $N=2$ ~\cite{Cheianov2005, Berkovich1987} by computing the Fredholm determinant. 
The momentum distribution in the spin-incoherent regime at $U\to \infty$ was then reconstructed numerically from the exact solution of the density matrix~\cite{Cheianov2005}. The power law singularity of the momentum distribution near $k=k_F$ in the Luttinger liquid regime disappears in the spin-incoherent regime due to the exponential decay of the Green's function~\cite{Cheianov2005}. The method by Cheianov et al~\cite{Cheianov2005} can be directly generalized to the spin-incoherent regime of the $SU(N)$ case at $U\to \infty$. However, due to the involved mathematics, we will leave it for a future study.

\section{Conclusions}\label{sec:summary}

In conclusion, we studied the one-dimensional $SU(N)$ Fermionic Hubbard model with $N$ spin species in the frame work of Luttinger liquid theory supplemented by the solutions of Bethe Ansatz equations. We focus on the metallic phase of the system, i.e., with  incommensurate fillings. 
The Green's function,  the momentum distribution and the tunneling density of states of the system are calculated in the Luttinger liquid regime $T\ll E_{{\rm spin}}$ and the Luttinger liquid parameter is determined from the Bethe Ansatz equations for arbitrary interaction. The theoretical results we obtained agree qualitatively with the recent experiments on one-dimensional $SU(N)$ Fermionic Hubbard model in alkaline earth atoms. 

We also studied the Green's function and tunneling density of states of spin-incoherent regime with $E_{{\rm spin}}\ll T\ll E_c$, which demonstrate significant difference from the Luttinger liquid regime. The Green's function decays exponentially with distance in the spin-incoherent regime instead of a power law in the Luttinger liquid regime and the tunneling density of state diverges at low energy in the former instead of vanishing in the latter. The momentum distribution in the spin-incoherent regime, however, remains a challenge in a general case of arbitrary interaction.

We compared the one-dimensional $SU(N)$ Fermionic Hubbard model at large $N$ with the one-dimensional spinless Bosoinic Hubbard model. Though the ground state energy of the Fermionic system at $N\to \infty$ is the same as the spinless Bosonic system with the same interaction and particle density, indicating Bosonic behaviors of the $SU(N)$ Fermionic system at large $N$, the single-particle Green's function and the momentum distribution of the two systems are not the same due to the different statistics they obey.

{\it Acknowledgement.} We acknowledge useful discussion with Hui Zhai and Zhenjie Li. This work is supported by the NSF of China under Grant No.11504195 (WC).

\

\

\


\end{document}